\documentclass{iau} 
\usepackage{graphicx}

\newcommand*{\Ledd}{L_{\rm Edd}}
\newcommand*{\calR}{{\cal R}}
\newcommand*{\calM}{{\cal M}}
\newcommand*{\calFg}{{\cal F}_{\rm g}}

\newcommand*{\Mdotbe}{\dot M_{\rm e}}

\newcommand*{\Mdott}{\dot M_{\rm t}}

\newcommand*{\lambdat}{\lambda_{\rm t}}

\newcommand*{\lambdacr}{\lambda_{\rm cr}}

\newcommand*{\xmin}{x_{\rm min}}

\newcommand*{\Mbh}{M_{\rm BH}}
\newcommand*{\Mg}{M_{\rm g}}
\newcommand*{\rhog}{\rho_{\rm g}}

\newcommand*{\cs}{c_{\rm s}}
\newcommand*{\rb}{r_{\rm B}}

\newcommand*{\rg}{r_{\rm g}}
\newcommand*{\rmin}{r_{\rm min}}

\newcommand*{\rhoinf}{\rho_{\infty}}
\newcommand*{\pinf}{p_{\infty}}
\newcommand*{\csinf}{c_{\infty}}
\newcommand*{\Tinf}{T_{\infty}}
\newcommand*{\rhotil}{\tilde\rho}
\newcommand*{\mpr}{m_{\rm p}}

\newcommand\rhos{\rho_*}

\newcommand\rhoDM{\rho_{\rm DM}}
\newcommand\rhon{\rho_{\rm n}}

\newcommand\rs{r_*}

\newcommand\Phig{\Phi_{\rm g}}

\newcommand\Psin{\Psi_{\rm n}}

\newcommand\Ms{M_{\star}}

\newcommand\MR{{\cal R}}
\newcommand\MRg{{\cal R}_{\rm g}}

\newcommand\sigv{\sigma_{\rm V}}

\newcommand\Tv{T_{\rm V}}

\newcommand\csig{\xi_{\rm g}}

\newcommand\betac{\beta_{\rm c}}

\title[Bondi accretion in two-component galaxies]
{Fully analytical solutions for Bondi accretion in galaxies with a
  central Black Hole}

\author[Luca Ciotti and Silvia Pellegrini]
{Luca Ciotti$^1$ \and
Silvia Pellegrini$^1$}

\affiliation{$^1$Department of Physics and Astronomy, University of Bologna, \\
via Gobetti 93/2, 40129 Bologna, Italy \\
email: {\tt luca.ciotti@unibo.it} \\[\affilskip]}

\pubyear{2018}
\volume{342}

\jname{Perseus in Sicily: from black hole to cluster outskirts}
\editors{K. Asada, E. de Gouveia dal Pino, H. Nagai, \\
R. Nemmen, M. Giroletti, eds.}

\begin{document}

\maketitle

\begin{abstract}

  The fully analytical solution for isothermal Bondi accretion on a
  black hole (MBH) at the center of JJ two-component Jaffe (1983)
  galaxy models is presented. In JJ models the stellar and total mass
  density distributions are described by the Jaffe profile, with
  different scale-lengths and masses, and to which a central MBH is
  added; all the relevant stellar dynamical properties can also be
  derived analytically. In these new accretion solutions the
  hydrodynamical and stellar dynamical properties are linked by
  imposing that the gas temperature is proportional to the virial
  temperature of the stellar component.  The formulae that are
  provided allow to evaluate all flow properties, and are then useful
  for estimates of the accretion radius and the mass flow rate when
  modeling accretion on MBHs at the center of galaxies.

\end{abstract}


\section{Introduction}

Observational and numerical investigations of accretion on massive
black holes (hereafter MBH) at the center of galaxies often lack the
resolution to follow gas transport down to the parsec scale. In these
cases, the {\it classical} Bondi (1952) solution on to an isolated
central point mass is then commonly adopted, for estimates of the
accretion radius (i.e., the sonic radius), and the mass accretion rate
(see Ciotti \& Pellegrini 2017, hereafter CP17, and references
therein).  However, two major problems affect the direct application
of the classical Bondi solution, namely the facts that 1) the boundary
values of density and temperature of the accreting gas are assigned at
infinity, and 2) in a galaxy, the gas experiences the gravitational
effects of the galaxy itself (stars plus dark matter), and the MBH
gravity becomes dominant only in the very central regions.  The
solution commonly adopted is to use values of the gas density and
temperature ``sufficiently near'' the MBH. It is therefore important
to quantify the systematic effects, on the estimates obtained from the
classical Bondi solution for the accretion radius and the mass
accretion rate, due to measurements taken at finite distance from the
MBH, and under the effects of the galaxy potential well. A first step
analysis of this problem was carried out in Korol et al. (2016,
hereafter KCP16) where the Bondi problem was generalized to the case
of mass accretion at the center of galaxies, including also the effect
of electron scattering on the accreting gas. CP17 showed that the
whole accretion solution can be given in an analytical way for the
{\it isothermal} accretion in Jaffe (1983) galaxies with a central
MBH.  Ciotti \& Pellegrini (2018, hereafter CP18, in preparation),
extend the study to JJ two-component galaxy models (Ciotti \& Ziaee
Lorzad 2018, hereafter CZ18), where the {\it stellar} and {\it total}
mass density distributions are both described by the Jaffe profile,
with different scale-lengths and masses, and a MBH is added at the
center.  In particular, JJ models offer the {\it unique} opportunity
to have a quite realistic family of galaxy models with a central MBH,
allowing both for the fully analytical solution of the Bondi
(isothermal) accretion problem, {\it and} the fully analytical
solution of the Jeans equations.

\section{The models}

\subsection{The Bondi solution}

In the Bondi problem, the gas is perfect, has a spatially infinite
distribution, and is accreting on to a MBH, of mass $\Mbh$. The gas
density $\rho$ and pressure $p$ are linked by the polytropic relation
\begin{equation} 
p = {k_{\rm B} \rho T\over <\mu>\mpr} = \pinf\rhotil^{\gamma},\quad
\rhotil\equiv{\rho\over\rhoinf},
\end{equation}
where $\gamma =1$ in the isothermal case, $\mpr$ is the proton mass,
and $\pinf$ and $\rhoinf$ are respectively the gas pressure and
density at infinity.  The sound speed is $\cs=\sqrt{\gamma p/\rho}$,
and in the isothermal case $T (r)=\Tinf$ and so $\cs (r)=\csinf$.
With the introduction of the Bondi radius $\rb$ and of the Mach number
$\calM$, given respectively by
\begin{equation} 
\rb \equiv {G\Mbh\over\csinf^2},\quad
\calM(r)={v(r)\over\cs(r)},
\end{equation}
the time-independent continuity equation becomes
\begin{equation}
x^2 \rhotil (x)\calM (x) ={\Mdott\over
  4\pi\rb^2\rhoinf\csinf}\equiv\lambda,
\quad x\equiv {r\over\rb},
\end{equation}
where $\Mdott$ is the mass accretion rate on the MBH at the center of
the galaxy, and $\lambda$ the accretion parameter. 

For a Jaffe galaxy of total mass $\Mg$ and scale-length $\rg$, the
gravitational potential and the two parameters determining the
accretion solution are (CP17):
\begin{equation} 
	\Phig={G\Mg\over\rg}\ln {r\over r+\rg},\quad
\calR \equiv {\Mg\over\Mbh},\quad 
\xi \equiv {\rg\over\rb}.
\end{equation}
In the Bondi problem for isothermal accretion reduces to the solution of
\begin{equation}
{\calM^2\over 2}-\ln\calM = f(x) -{\rm e}^{\lambda},\quad
f = {\chi\over x}  - {\calR\over\xi}\ln {x\over x+\xi} +  2\ln  x,
\end{equation}
where $\chi\equiv 1- L/\Ledd$ measures the effect of electron
scattering radiation pressure due to the accretion luminosity
$L$. Solutions exist only for $\lambda\leq\lambdat$, the {\it
  critical} accretion parameter, a value determined by the position of
the minimum $\xmin$ for the function $f$.  For the Jaffe galaxy the
position of the only minimum of $f$ (corresponding to the sonic radius
of the critical solution) can be calculated analytically, with
$\xmin = \xmin(\chi,\calR,\xi)$, so that quite surprisingly one can
evaluate $\lambdat$ analytically.  In the peculiar case of $\chi =0$
(and/or $\Mbh =0$), a solution of the accretion problem is possible
only for $\calR\geq 2\xi$. Moreover, the radial trend of the Mach
number can also be calculated analytically in terms of the so-called
$W$ Lambert-Euler function.

\begin{figure*}
\includegraphics[height=0.5\textwidth, width=0.5\textwidth]{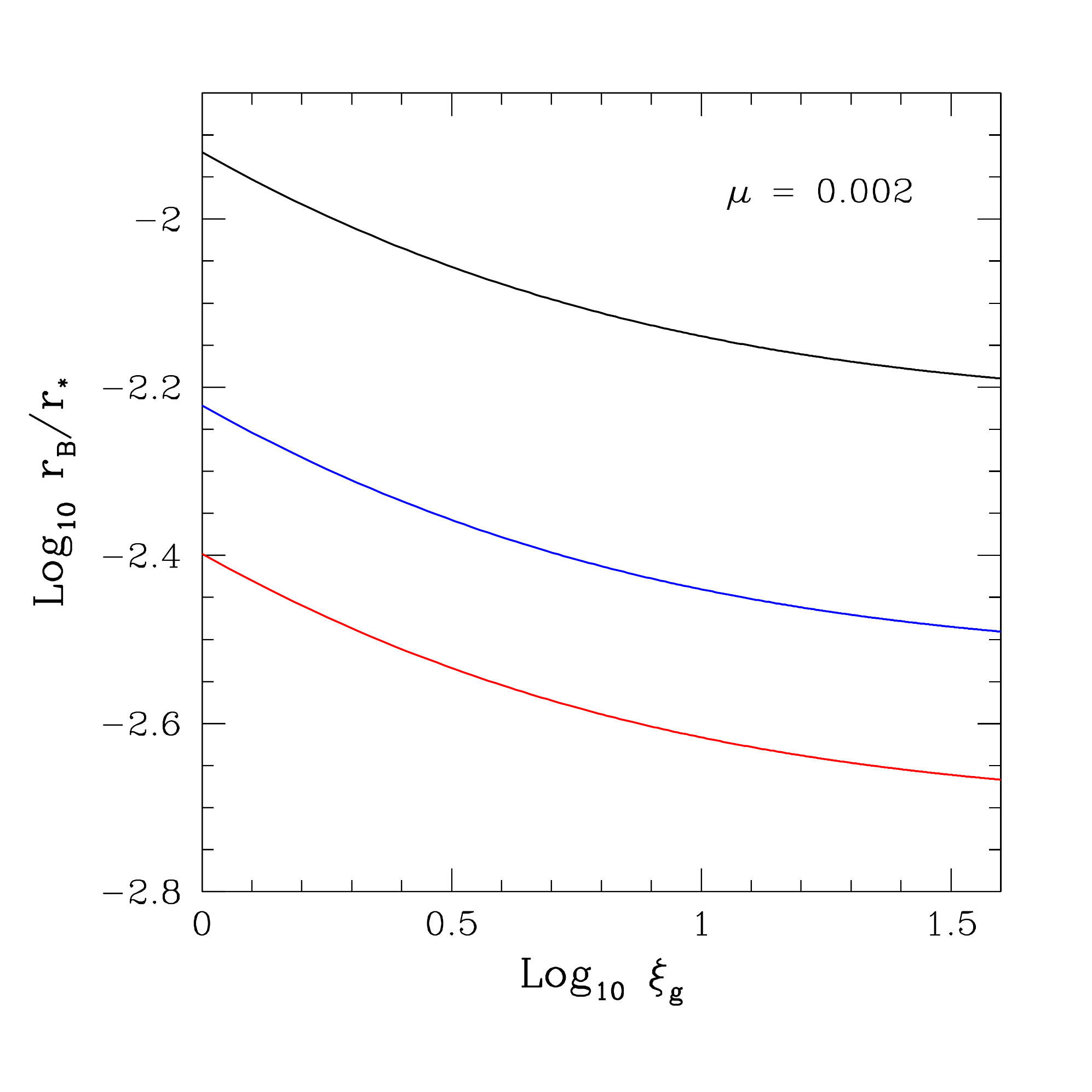}
\includegraphics[height=0.5\textwidth, width=0.5\textwidth]{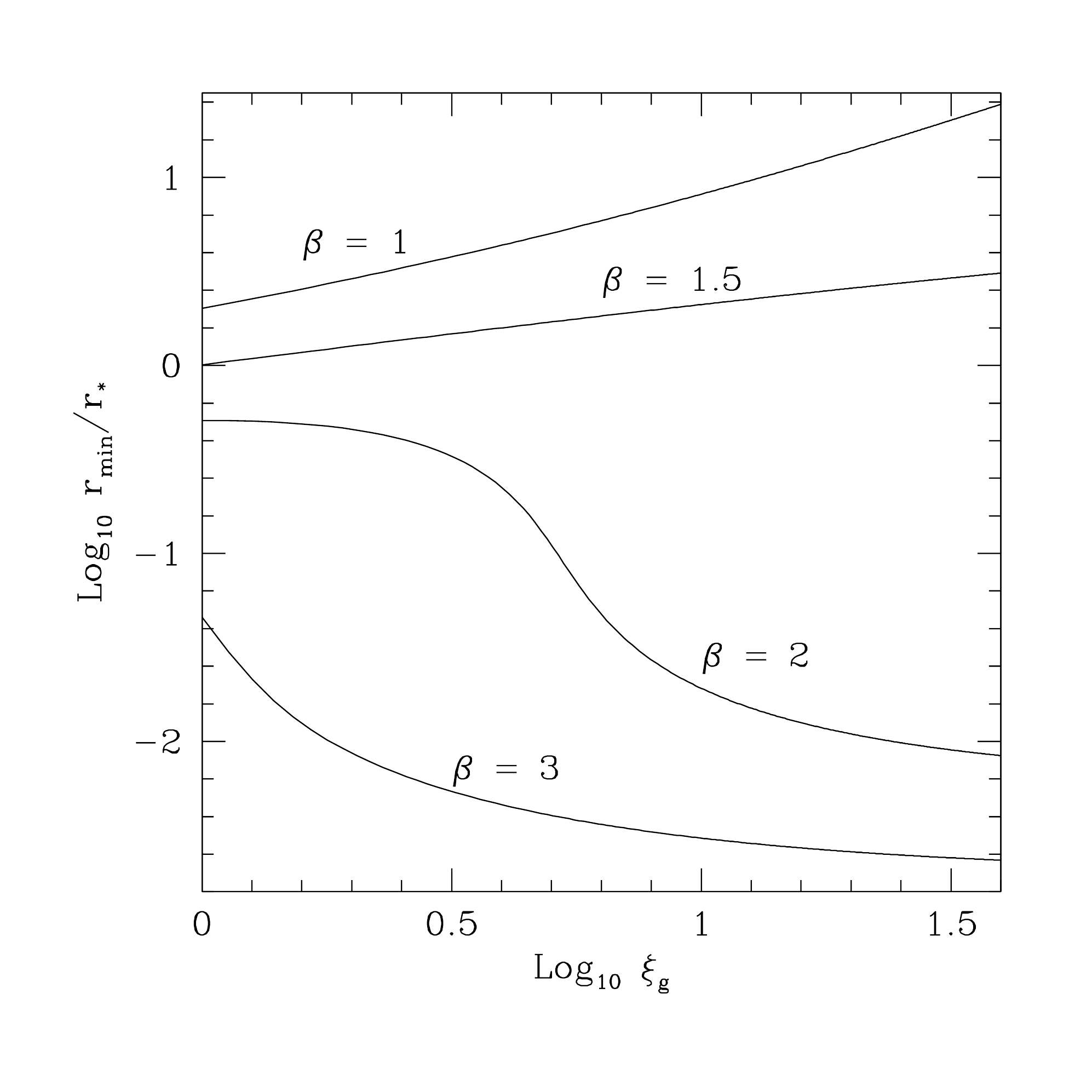}
\vskip -2truecm 
\includegraphics[height=0.5\textwidth, width=0.5\textwidth]{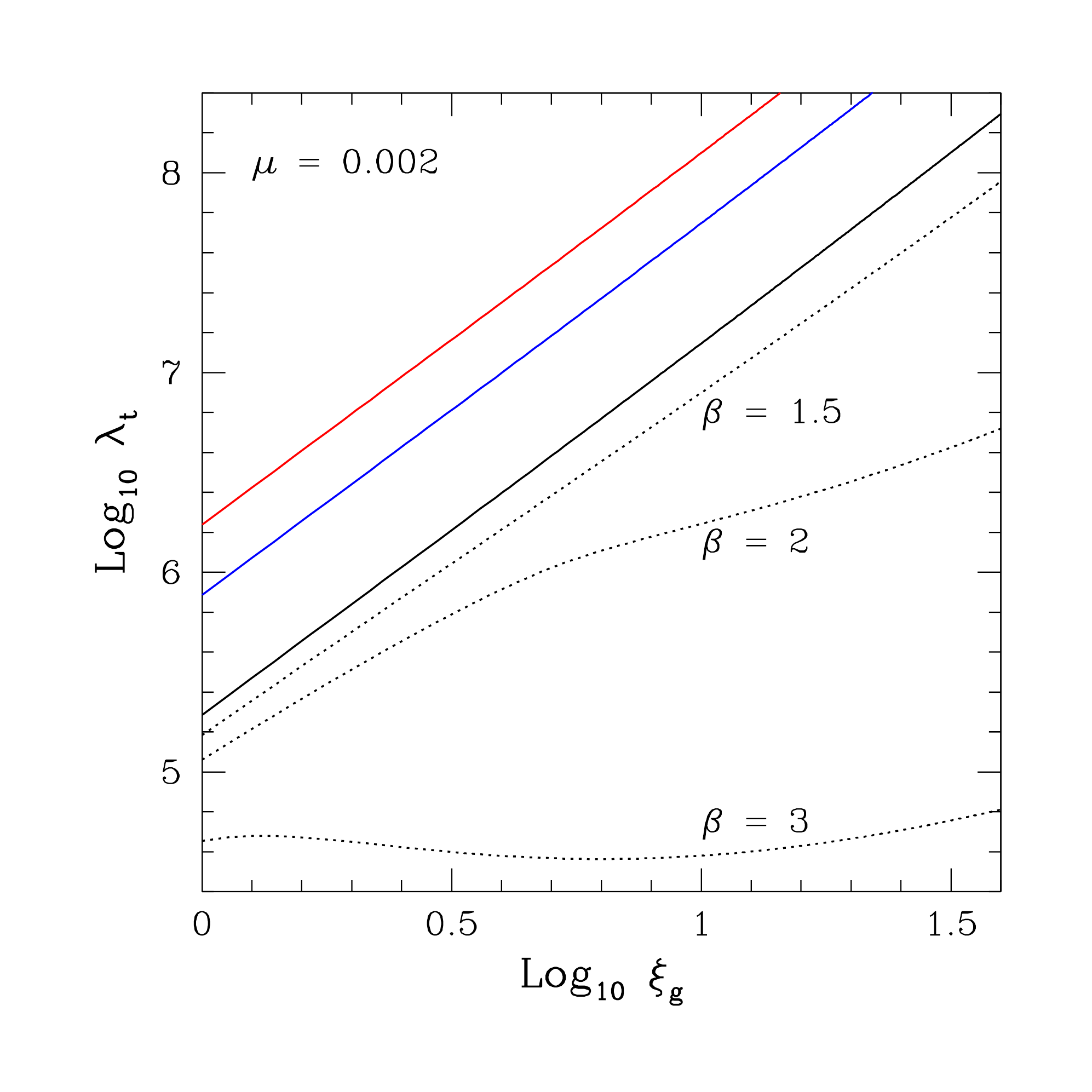}
\hskip -0.3truecm 
\includegraphics[height=0.59\textwidth, width=0.59\textwidth]{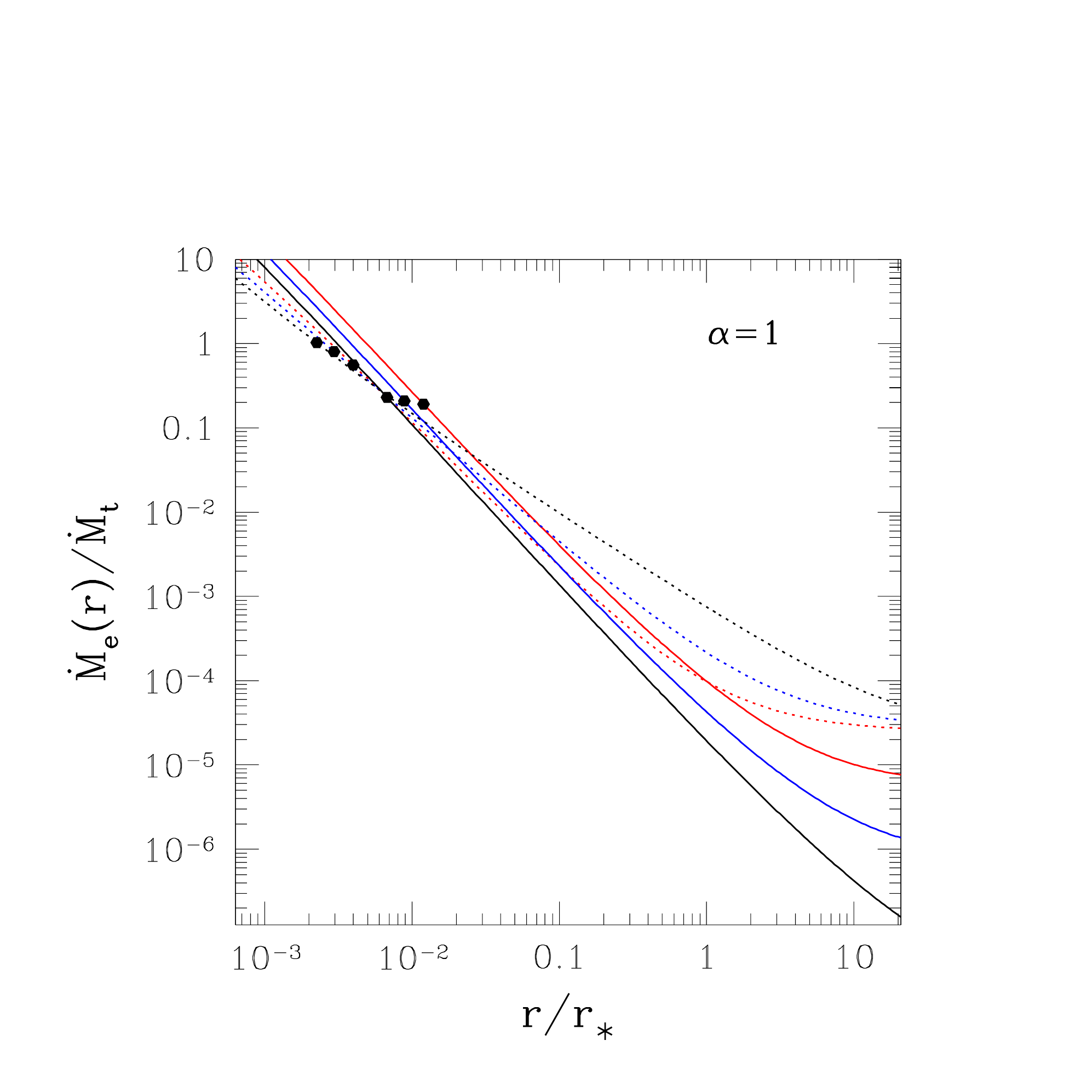}
\caption{Relevant scale-lengths of the isothermal accretion solution
  in JJ models, as a function of $\csig=\rg/\rs$. A MBH-to-galaxy
  stellar mass ratio $\mu=2\times 10^{-3}$ is assumed, and
  $\Tinf=\beta\Tv$.  Top left: the ratio $\rb/\rs$ with $\beta=1$ and
  $\alpha =1$ (black), 2 (blue), 3 (red).  Top right: ratio
  $\rmin/\rs$ for $\alpha =1$ and $\beta =1,3/2,2,3$: for large values
  of $\MR$ and $\beta <\betac$ the ratio is almost independent of
  $\alpha$ but strongly dependent on gas temperature; for
  $\beta >\betac$, the sonic radius collapses near the center.  Bottom
  left: the critical accretion parameter $\lambdat$ as a function of
  $\csig$, for $\alpha=1,2,3$, $\chi =1$, and $\beta=1$ (solid
  lines). The dotted curves refer to $\alpha =1$ and three different
  values of $\beta$.  Bottom right: ratio between the estimate of the
  accretion rate $\Mdotbe$ and the true accretion rate $\Mdott$, as a
  function of $r/\rs$, in the minimun halo case ($\alpha =1$), and
  $\csig=1$ (red), $\csig=3$ (blue), and $\csig=20$ (black). The
  dotted lines correspond to $\beta =3$, and the solid dots mark the
  position of the Bondi radius.}
\label{}
\end{figure*}

CP18 apply the previous results to JJ models, that are characterized
by a {\it total} Jaffe density distribution $\rhog$ (stars plus dark
matter) of total mass $\Mg$ and scale-length $\rg$, and a stellar
Jaffe distribution of stellar mass $\Ms$, and scale radius $\rs$.
Remarkably, almost all the stellar dynamical properties of JJ models
with a central MBH can be expressed by analytical functions (CZ18), so
that they are a family of two-component galaxy models with a central
MBH allowing for a full analytical treatment of Bondi accretion and of
stellar dynamics (CP18).

\subsection{Structure and dynamics of the JJ models}

The stellar and total density distributions of JJ galaxies are given
respectively by 
\begin{equation}
\rhos(r)=
{\rhon\over s^2(1+s)^2},\quad  
\rhog(r)= {\rhon\MRg\csig\over s^2(\csig+s)^2},\quad
s\equiv {r\over\rs},
\end{equation}
where we define
\begin{equation}
\rhon\equiv{\Ms\over 4\pi\rs^3},\quad 
\Psin\equiv{G\Ms\over\rs},\quad 
\mu\equiv{\Mbh\over\Ms},\quad
\MRg\equiv 
{\Mg\over\Ms},\quad \csig \equiv {\rg\over\rs}.
\end{equation} 
Typically, $\mu\approx 10^{-3}$. When $\csig\geq 1$, the density
distribution of the dark halo $\rhoDM =\rhog -\rhos$ is nowhere
negative, provided that $\MRg=\alpha\csig$ with $\alpha\geq 1$.  A
galaxy with $\alpha =1$ is called a {\it minimum halo} model, and the
associated halo is well approximated by the NFW (Navarro et al. 1997)
profile over a very large radial range (CZ18). For JJ models the
Osipkov-Merrit anisotropic Jeans equation, and the projected values of
the velocity dispersion at the galaxy center, can be expressed
analytically. In particular, the virial temperature of the stellar
component can be written as $\Tv=<\mu>\mpr\sigv^2/3$, where the virial
velocity dispersion of stars is given by
$\sigv^2=\Psin\alpha\calFg(\csig)$, and $\calFg(\csig)$ is a simple
analytical function, with $\calFg (1)=1/2$ and $\calFg(\infty)=1$.

\subsection{Linking stellar dynamics to fluidodynamics}

In CP18 the idea is to self-consistently ``close'' the accretion
solution, determining a fiducial value for the gas temperature as a
function of the galaxy model hosting accretion.  For
assigned values of $\csig\geq 1$, $\MRg$ (or $\alpha$), and $\mu$, we
fix $\Tinf=\beta\Tv$, with $\beta >0$, so that
$\csinf=\sigv\sqrt{\beta/3}$. All the accretion parameters can
therefore be obtained in terms of the galaxy properties as
\begin{equation}
\MR= {\alpha\csig\over\mu},\quad
\xi ={\MR\beta\calFg\over 3},\quad
{\rb\over\rs}={3\mu\over \alpha\beta\calFg},\quad
{\rmin\over\rs}=\xmin(\chi,\MR,\xi) \, {\rb\over\rs},
\end{equation}
where $\rmin$ is the position of the sonic radius. Then the critical
accretion parameter $\lambdat$, and the Mach number profile, can be
computed analytically using the results of CP17. It turns out that for
JJ models with $\chi =0$ (and/or $\Mbh =0$) the Bondi solution exists
only for $\beta \leq \betac\equiv 3/(2\calFg)$ with
$3/2\leq\betac\leq 3$, and that $\betac$ determines the behavior of
the solution also in presence of a MBH. The first three panels in
Fig. 1 show some representative cases of the quantities in
eq. (2.8). One of the most relevant features is the considerable jump
of $\rmin$ from very external to very internal galactic regions, even
for a slight increase of the gas temperature. Finally, CP18 evaluate
the departure of the estimated mass accretion rate
$\Mdotbe(r)\equiv 4 \pi \rb^2 \lambdacr \rho(r)\csinf$ obtained from
the classical Bondi solution, from the true value $\Mdott$, as a
function of the distance from the center:
\begin{equation}
{\Mdotbe(r)\over\Mdott}={\lambdacr\rhotil(x)\over\lambdat}={\lambdacr\over
  x^2\calM(x)},
\end{equation}
where $\calM(x)$ is the solution of eq. (2.5).  Here $\rho (r) $ is
taken along the solution for accretion within the potential of the
galaxy and used as ``proxy'' for the true value $\rhoinf$, and
$\lambdacr = {e\rm}^{3/2}/4$ is the isothermal critical accretion
parameter of the Bondi solution on an isolated MBH. Note how the bias
increases from values much lower than unity in the outer galactic
regions to very large values near the center; this allows to directly
estimate the so-called ``boost factor'' (see CP18 for a thorough
discussion).

\end{document}